\documentclass[conference]{IEEEtran}
\IEEEoverridecommandlockouts
\usepackage{cite}
\usepackage{amsmath,amssymb,amsfonts}
\usepackage{algorithmic}
\usepackage{graphicx}
\usepackage{textcomp}
\usepackage{xcolor}
\usepackage[T1]{fontenc}
\usepackage{times}

\def\BibTeX{{\rm B\kern-.05em{\sc i\kern-.025em b}\kern-.08em
    T\kern-.1667em\lower.7ex\hbox{E}\kern-.125emX}}
\usepackage{listings}
\usepackage{xcolor}

\IEEEpubid{\begin{minipage}{\textwidth}
\vspace{40pt}
\centering\scriptsize
\textcopyright~2026 IEEE. Personal use of this material is permitted. Permission from IEEE must be obtained for all other uses, in any current or future media, including reprinting/republishing this material for advertising or promotional purposes, creating new collective works, for resale or redistribution to servers or lists, or reuse of any copyrighted component of this work in other works. 
\end{minipage}}

\begin{document}

\title{Compiler-Guided Inference-Time Adaptation: Improving GPT-5 Programming Performance in Idris}

\author{
\IEEEauthorblockN{Minda Li and Bhaskar Krishnamachari}
\IEEEauthorblockA{\textit{Ming Hsieh Department of Electrical and Computer Engineering} \\
\textit{Viterbi School of Engineering, University of Southern California}\\
Los Angeles, California \\
\{mcli, bkrishna\}@usc.edu}

}

\maketitle
\IEEEpubidadjcol

\begin{abstract}
GPT-5, a state of the art LLM from OpenAI, demonstrates strong performance in widely used programming languages such as Python, C++, and Java; however, its ability to operate in low-resource or less commonly used languages remains underexplored. This work investigates whether GPT-5 can effectively acquire proficiency in an unfamiliar functional programming language, Idris, through iterative, feedback-driven prompting. We first establish a baseline showing that with zero-shot prompting the model solves only 22 out of 56 Idris exercises using the platform Exercism, substantially underperforming relative to higher-resource languages (45/50 in Python and 35/47 in Erlang). We then evaluate several refinement strategies, including iterative prompting based on platform feedback, augmenting prompts with documentation and error classification guides, and iterative prompting using local compilation errors and failed test cases. Among these approaches, incorporating local compilation errors yields the most substantial improvements. Using this structured, error-guided refinement loop, GPT-5’s performance increased to an impressive 54 solved problems out of 56. These results suggest that while large language models may initially struggle in low-resource settings, structured compiler-level feedback can play a critical role in unlocking their capabilities.
\end{abstract}

\section{Introduction}

Large language models such as OpenAI's GPT-5 model exhibit strong performance across widely used programming languages, including Python, C++, and Java. Their high proficiency is supported by vast amounts of training data and documentation available online. However, far less is known about how effectively these models can work with low-resource or less commonly represented languages. Evaluating performance in such languages offers insight into how well LLMs generalize beyond high-frequency training examples and whether they can be taught new abstractions through iterative prompting.

Idris serves as an ideal test case for this question. As a dependently typed functional programming language, it is significantly underrepresented in public code repositories compared to mainstream languages. GitHub repository counts illustrate this disparity: Idris has 2,275 repositories, similar in scale to other niche languages such as Forth (2,255) and Agda (3,049). In contrast, higher-level functional languages like Racket (22,655), industrial languages like Erlang (33,058) and Elixir (121,841), and especially general-purpose languages like Python (24,286,135) and C++ (6,182,683) have vastly greater representation. This imbalance suggests that LLMs are likely to have far less exposure to Idris during training. Thus Idris provides a controlled setting to test whether models can adapt to languages they have not been deeply trained on.

To contextualize performance on Idris, we also established model baselines in more common languages. Because Exercism language tracks differ in size, cross-language baselines were computed over the largest overlapping subsets of identical exercises: all 56 available Idris problems, 50 problems shared between the Python and Idris tracks, and 47 problems shared between the Erlang and Idris tracks. In Python, GPT-5 solved 90\% of problems (45/50), successfully compiling all solutions but failing a small portion of test cases. In Erlang, a less common but still significantly more represented language when compared to Idris, the model solved 74\% of problems (35/47). These baselines provide useful comparative anchors for evaluating how the model behaves when confronted with increasingly low-resource languages.

Due to its relative rarity, strict type system, and unfamiliar functional patterns, Idris offers a unique environment to explore a key research question: Can GPT-5 learn a new programming language through iterative, feedback-driven refinement?

This study addresses the following research questions:

\begin{itemize}
    \item \textbf{RQ1:} How does GPT-5 perform in a low-resource, dependently typed programming language compared to higher-resource programming languages?
    \item \textbf{RQ2:} To what extent can iterative, error-guided prompting, particularly when incorporating compiler-level feedback, improve GPT-5’s performance in a low-resource programming language?
\end{itemize}

In summary, this paper makes three contributions. First, we provide (to our knowledge) the first measurement study of GPT-5’s performance on the Idris Exercism track under a clearly specified evaluation protocol. Second, we compare four inference-time refinement strategies, ranging from zero-shot prompting to retrieval-augmented prompting and iterative compile/test feedback loops, and quantify their impact on solve rates. Third, our results suggest that compiler diagnostics are the dominant learning signal driving improvements, as evidenced by the attempt breakdowns and observed error patterns.

%https://news.ycombinator.com/item?id=42620047

\section{Related Work}

Our study situates itself at the intersection of evaluating Large Language Models (LLMs) on low-resource programming languages and enhancing their performance through iterative, compiler-guided refinement.

\subsection{Multilingual Evaluation and Low-Resource Adaptation}
While early benchmarks for code generation, such as HumanEval, focused primarily on Python \cite{chen2021codex}, recent work has highlighted the performance disparity across programming languages. \textit{MultiPL-E} extended unit-test-based evaluation to many languages, confirming that model performance correlates strongly with language representation in the training corpus \cite{cassano2022multiplE}.
To address this data scarcity, \textit{MultiPL-T} demonstrated that training-time transfer from high-resource languages—using synthesized, test-validated data—can improve proficiency in low-resource settings \cite{cassano2023multiplT}.
In contrast to these training-centric approaches, our work investigates \textit{inference-time} adaptation, exploring whether a general-purpose model can acquire proficiency in a niche language like Idris solely through iterative feedback and documentation.

\subsection{Iterative Refinement and Feedback Loops}
Recent research emphasizes moving beyond single-shot generation toward iterative feedback loops. \textit{Self-Refine} demonstrated that LLMs can improve their own outputs through iterative critique and revision without additional training \cite{madaan2023selfrefine}. Specific to code, \textit{SELF-DEBUGGING} enables models to execute code, analyze execution traces, and generate feedback to fix errors \cite{chen2024selfdebugging}. Frameworks like \textit{LLMLOOP} further automate this process by closing the loop between generation, compilation, and testing \cite{ravi2025llmloop}. Our study leverages these iterative principles but distinguishes itself by prioritizing \textit{compiler-level} feedback, which we find to be the critical bottleneck in dependently typed languages compared to execution-based feedback.

\subsection{Compiler Diagnostics and Formal Constraints}
The use of compiler diagnostics for program repair is well-established. Pre-LLM systems like \textit{DrRepair} utilized graph-based modeling to link source code symbols with compiler error messages for automated fixing \cite{yasunaga2020drrepair}.
In the era of LLMs, this approach has extended to formal environments. \textit{Baldur} generates proofs and repairs failures using error context from proof assistants, mirroring the use of compiler feedback for repair \cite{first2023baldur}, while \textit{LeanDojo} provides a retrieval-augmented environment for theorem proving \cite{yang2023leandojo}. Furthermore, Blinn et al.\ showed that exposing static semantics, such as typed holes, can effectively constrain and guide LLM generation \cite{blinn2024typedholes}. Our results corroborate these findings, demonstrating that in strict functional settings, structured compiler feedback provides a more potent learning signal than natural language prompting alone.

\section{Methodology}

All code for the following methodologies is available at [a URL that will be provided upon acceptance, omitted for double-blind review].
%, https://github.com/ANRGUSC/Idris-gpt/tree/main. 
When this project first began, we were interested in finding programming languages that were rarely used and unfamiliar. Through this search, we found the platform Exercism. Exercism has a variety of languages ranging from Python to Cobol, encompassing 78 different languages. We narrowed in on Idris, a dependently typed program. Idris had 56 problems (excluding the start Hello, World! problem). Exercism had an option to download each problem locally to your computer, so we utilized this feature.

After all 56 problems were downloaded locally onto our computer, we utilized two different python scripts to establish a baseline. The first python script was a query writer. To automate query writing for all of the problems, the python script first extracted the README.md content that contained the problem description, and also scraped the starter lines of code provided in the idris file. The query writer format was then written as following:

Below is a sample query produced.
\begin{lstlisting}[label={lst:query}]
Write a Idris program to do the following, and put this program in a json object as the value corresponding to the key 'code'.
# Accumulate

Welcome to Accumulate on Exercism's Idris Track.
If you need help running the tests or submitting your code, check out `HELP.md`.

## Instructions

Implement the `accumulate` operation, which, given a collection and an operation to perform on each element of the collection, returns a new collection containing the result of applying that operation to each element of the input collection.

Given the collection of numbers:

- 1, 2, 3, 4, 5

And the operation:

- square a number (`x => x * x`)

Your code should be able to produce the collection of squares:

- 1, 4, 9, 16, 25

Check out the test suite to see the expected function signature.

## Restrictions

Keep your hands off that collect/map/fmap/whatchamacallit functionality provided by your standard library!
Solve this one yourself using other basic tools instead.

## Source

### Contributed to by

- @keiravillekode
- @ktosiek
- @kytrinyx
- @NobbZ
- @yurrriq

### Based on

Conversation with James Edward Gray II - http://graysoftinc.com/
Write code that fits the following template. The program should continue below the following lines:
module Accumulate

export
accumulate : (a -> b) -> List a -> List b
accumulate f [] = []
accumulate f (x :: xs) = f x :: accumulate f xs
\end{lstlisting}
Once all of the queries were generated, a second Python script was written to send each query to GPT-5 through an OpenAI API key and automatically save whatever code the model produced. The script read each query file, sent its contents as a request to the GPT-5 model, and then captured the model’s response. Because every query asked GPT-5 to return its answer in JSON format with the Idris code stored under a "code" key, the script simply extracted that field from the response and saved it to the file that provided the starter code. This process allowed us to generate and store all model outputs consistently. All answers were submitted back to Exercism, and we noted whether the problem was passed or not. 

After the baseline was established on the number of problems that could be solved by GPT-5, we then began to investigate how to improve its performance rate through a few different methods. First, we evaluated iterative prompting using failed test cases returned by the Exercism platform, considering both a single iteration and five iterations of feedback as a single refinement approach. Next, we explored prompt augmentation strategies by supplying additional external information for problems not initially solved. These included (i) a constructed error documentation manual derived from common baseline errors and (ii) the official Idris reference documentation, each provided alongside the problem query. Finally, we evaluated an extensive iterative refinement strategy consisting of up to 20 iterations of local compilation and analysis of failed test cases, where compiler diagnostics and test failures were incorporated directly into subsequent prompts. 

\subsection{Method 1: Incorporating Exercism Error Messages into the Query}

After submitting each solution to the Exercism platform, any unsolved problems returned error messages. We incorporated these error messages directly into subsequent queries, instructing GPT-5 to correct the specific issues identified by the failing tests. This approach successfully resolved a small number of problems, motivating us to extend the procedure by applying up to five iterative rounds of error-guided refinement for each unsolved problem.

\subsection{Method 2: Error Avoidance Manual}

While reviewing the initial error messages, several recurring patterns emerged. We documented all errors from the first testing round and compiled them into a single dataset. We then asked GPT-5 to load this dataset into context and classify the errors into meaningful categories—such as syntax errors, missing libraries, unfilled holes, and logical mistakes. For each category, GPT-5 generated a description, identified how frequently it occurred, and explained why these errors arise in Idris. Using this classification, we prompted GPT-5 to create a reference manual: a document outlining common Idris pitfalls, how to avoid them, and best practices for writing valid Idris code. The goal was to determine whether providing GPT-5 with this self-generated manual would reduce repeated mistakes and improve its ability to produce correct Idris solutions.

To test whether a reference manual could improve GPT-5’s Idris performance, we stored the manual in a vector store so the model could retrieve relevant sections during each query. The script first uploads the PDF into a vector store, where it is embedded, chunked, and indexed for semantic search. When solving a new Idris problem, the model retrieves the most relevant passages from this store and uses them as context before generating code. This allowed GPT-5 to consistently reuse the same knowledge without re-uploading the document and reduced repeated errors across tasks.

\subsection{Method 3: Idris Reference Manual}

In addition to the self-generated error avoidance manual, we evaluated whether providing GPT-5 with official language documentation would further improve its performance on Idris exercises. We identified an external Idris reference manual in PDF form that comprehensively describes the language’s syntax, type system, standard library usage, and common programming patterns. This document was uploaded to a vector store and embedded to enable semantic retrieval.

Similar to Method 2, the reference manual was chunked and indexed so that relevant sections could be retrieved dynamically during problem solving. For each unsolved Idris exercise, the model first queried the vector store to retrieve documentation passages most relevant to the current problem and then incorporated this retrieved context into the prompt before generating a solution.

By comparing this method against the error avoidance manual, we aimed to assess whether access to official language documentation alone is sufficient to improve performance, or whether model-generated, error-specific guidance provides greater benefit for solving low-resource language tasks.

\subsection{Method 4: Local Compilation and Failed Tests}

In the final and most extensive refinement strategy, we evaluated an iterative, compiler-driven approach that incorporated both local compilation errors and failed test cases into the prompting loop. Unlike Methods 1–3, which relied primarily on platform-reported (Exercism) test failures or static reference materials, this method leverages detailed compiler diagnostics generated during local builds of each Idris solution.

For each unsolved exercise, GPT-5 first generated an initial solution, which was then compiled locally using the Idris compiler. If compilation failed, the resulting compiler error messages—including type errors, missing definitions, pattern-matching failures, or totality violations—were captured and incorporated directly into the subsequent prompt. GPT-5 was instructed to revise the code specifically to address the reported compilation issues. This process was repeated iteratively until the program successfully compiled or the predefined iteration limit of 20 was reached. For each problem, the refinement loop terminated immediately upon successful compilation and passing of all local tests, at which point the solution was submitted to Exercism; otherwise, the loop was halted after a maximum of 20 iterations. 

Once a solution compiled successfully, it was then executed against the full local test. Any failing test cases produced additional diagnostic output, which was likewise appended to the next query and used to guide further revisions. Only after a solution both compiled successfully and passed all local test cases was it submitted to the Exercism platform. This approach ensured that submissions were syntactically valid, type-correct, and functionally correct prior to platform evaluation.

This method extends the error-guided prompting strategy explored in Method 1 by incorporating lower-level compiler feedback rather than relying solely on high-level test failures. By exposing GPT-5 to precise compiler diagnostics and enforcing correctness at each stage of the development pipeline, this approach provided substantially stronger guidance for correcting errors.

\section{Results and Analysis}
We first present a qualitative example illustrating compiler-guided refinement in practice, followed by quantitative results summarizing performance gains across all Idris exercises.

To illustrate compiler-guided inference-time adaptation, we present a concise case study from the Bob exercise. The initial zero-shot solution produced by GPT-5 appeared semantically reasonable but failed to compile due to incorrect assumptions about Idris syntax and available standard library functions.
Initial zero-shot output (excerpt):
\begin{lstlisting}[label={lst:bob}]
module Bob

export
response : String -> String
response remark
  | isSilence remark = "Fine. Be that way!"
  | isYellQuestion remark = "Calm down, I know what I'm doing!"
  | isQuestion remark = "Sure."
  | isYell remark = "Whoa, chill out!"
  | otherwise = "Whatever."

\end{lstlisting}
The initial zero-shot solution failed due to syntax errors, missing imports, and undefined helper functions—common failure modes in Table II. Through iterative refinement, GPT-5 restructured control flow, removed unsupported imports, and explicitly defined character-level predicates.
Final converged solution (excerpt):
\begin{lstlisting}[label={lst:bob2}]
module Bob

import Data.String
import Data.List

export
response : String -> String
response remark =
  let t = trim remark in
    if isEmpty t then "Fine. Be that way!"
    else if isYell t && isQuestion t then "Calm down, I know what I'm doing!"
    else if isYell t then "Whoa, chill out!"
    else if isQuestion t then "Sure."
    else "Whatever."
\end{lstlisting}
After five iterations, the solution compiled successfully and passed all Exercism test cases. This example demonstrates how compiler diagnostics provide precise, instance-specific feedback that enables LLMs to adapt their outputs to the strict semantics of low-resource, statically typed languages.

\begin{table}[h]
\centering
\caption{Performance Gains from Iterative Refinement Methods (n = 56).}
\label{tab:iterative_results}
\footnotesize
\setlength{\tabcolsep}{4pt}
\resizebox{\columnwidth}{!}{%
\begin{tabular}{|l|c|c|c|}
\hline
\textbf{Method} & \textbf{Solved} & \textbf{+Solved} & \textbf{\% Solved} \\
\hline
Baseline (GPT-5.0) & 22 & -- & 39 \\
\hline
1 Iteration with Failed Test Cases & 27 & +4 & 48 \\
\hline
5 Iterations with Failed Test Cases & 27 & +4 & 48 \\
\hline
Error Documentation Manual & 30 & +7 & 54 \\
\hline
Idris Reference Manual & 34 & +11 & 61 \\
\hline
20 Iterations of Compiling and Failed Tests & 54 & +31 & 96 \\
\hline
\end{tabular}}
\end{table}
Table~\ref{tab:iterative_results} summarizes the performance gains achieved by each iterative refinement method across the 56 Idris exercises on the platform Exercism. The baseline GPT-5 model solved\footnote{By solved, we mean passing all test cases on Exercism.} 22 problems (39\%), leaving 33 unsolved problems to which subsequent refinement methods were applied (independently, not sequentially). Incorporating failed test case feedback from the Exercism platform yielded modest improvements: both one and five iterations of error-guided prompting increased the number of solved problems to 27 (+4), corresponding to a 48\% success rate, with no additional gains observed beyond a single iteration.
Providing supplemental reference materials produced larger improvements. Supplying a manually constructed error documentation manual increased performance to 30 solved problems (54\%), while augmenting prompts with the official Idris reference manual further improved performance to 34 solved problems (61\%). The largest performance gains were achieved using the local compilation and failed-test refinement strategy. By iteratively incorporating compiler diagnostics and local test failures into the prompting loop, GPT-5 solved 54 out of 56 problems (96\%), substantially outperforming all other methods.
Two problems remained unsolved after the 20-iteration limit. In both cases, repeated automated fixes introduced cascading compilation issues in helper definitions, types, and the test harness, preventing stable convergence despite multiple attempts.

\includegraphics[width=\columnwidth]{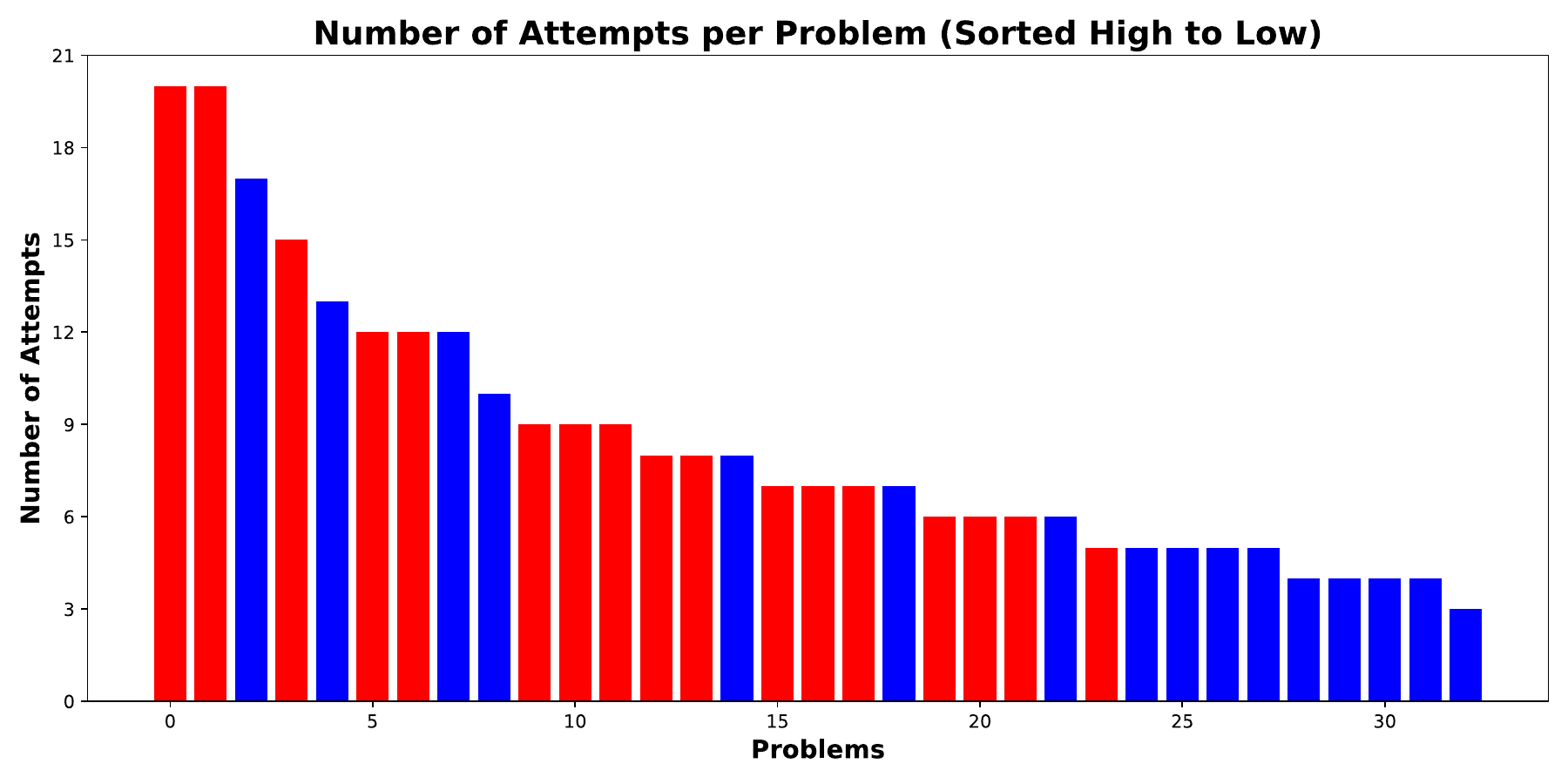}
The bar chart above shows the 33 problems that remained unsolved after the baseline evaluation, sorted from the highest to the lowest number of attempts required to solve the problem by GPT-5 using Method 4. Bars shown in blue represent problems that were successfully solved using at least one of Methods 1–3 (platform-based test feedback, the error documentation manual, or the Idris reference manual). Bars shown in red correspond to problems that were solved exclusively using Method 4, and they were not solved by method 1-3. This visualization highlights that problems requiring the greatest number of attempts were predominantly solvable only through compiler-driven refinement, underscoring the effectiveness of Method 4 to address more challenging cases.
\includegraphics[width=\columnwidth]{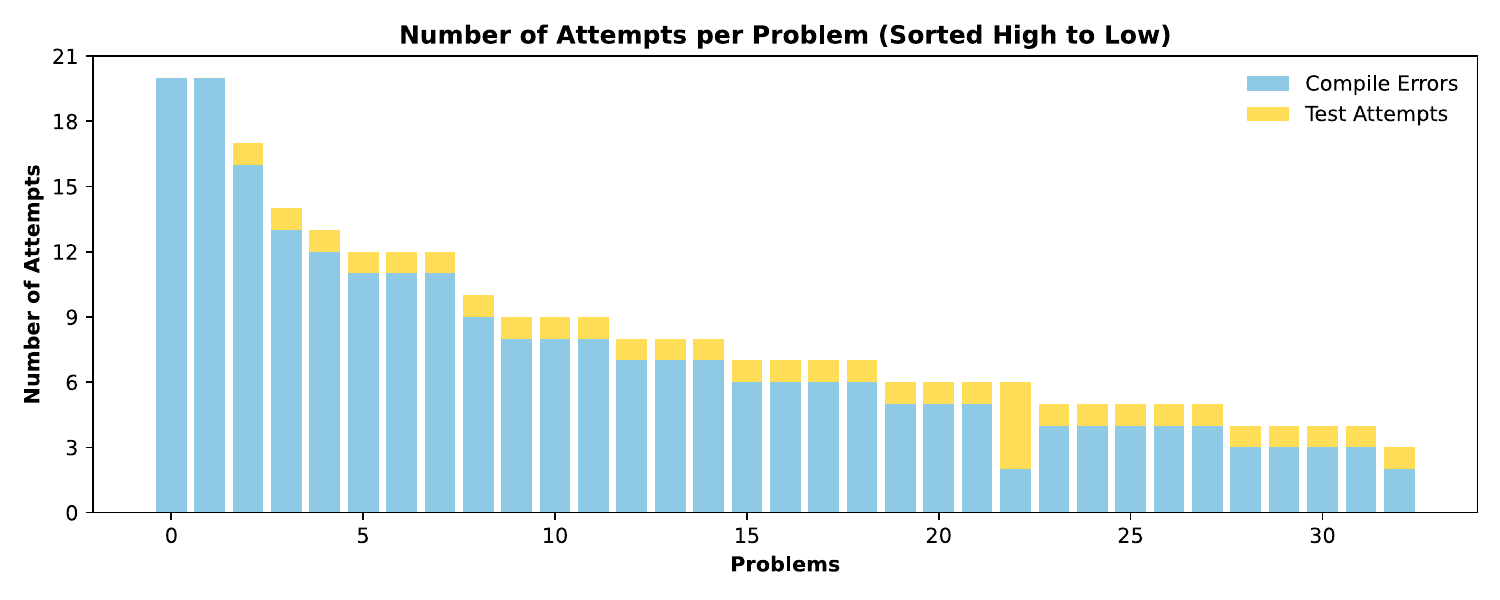}
The above bar chart illustrates the number of attempts required by Method 4 for each of the 33 initially unsolved problems, decomposed into compiler-driven fixes and test-driven fixes. The results show that the majority of attempts for difficult problems are dominated by compiler corrections rather than test failures. For many problems, a substantial number of iterations were required to resolve compiler errors. In contrast, once compilation succeeded, relatively few additional iterations were typically needed to address failing test cases.
This pattern suggests that, in the Idris setting, the primary bottleneck for large language models lies in satisfying the language’s strict type system and compilation requirements rather than in correcting logical errors exposed by tests. Problems requiring the highest number of total attempts tend to exhibit a long sequence of compiler fixes followed by a short test-refinement phase, indicating that compiler-level feedback provides the most informative signal for guiding model corrections. Overall, the breakdown highlights why Method 4 substantially outperforms other refinement strategies: by explicitly iterating on compiler diagnostics, the model is able to resolve low-level semantic and type errors that are not captured by test failures alone, enabling successful convergence on even the most challenging problems.
\includegraphics[width=\columnwidth]{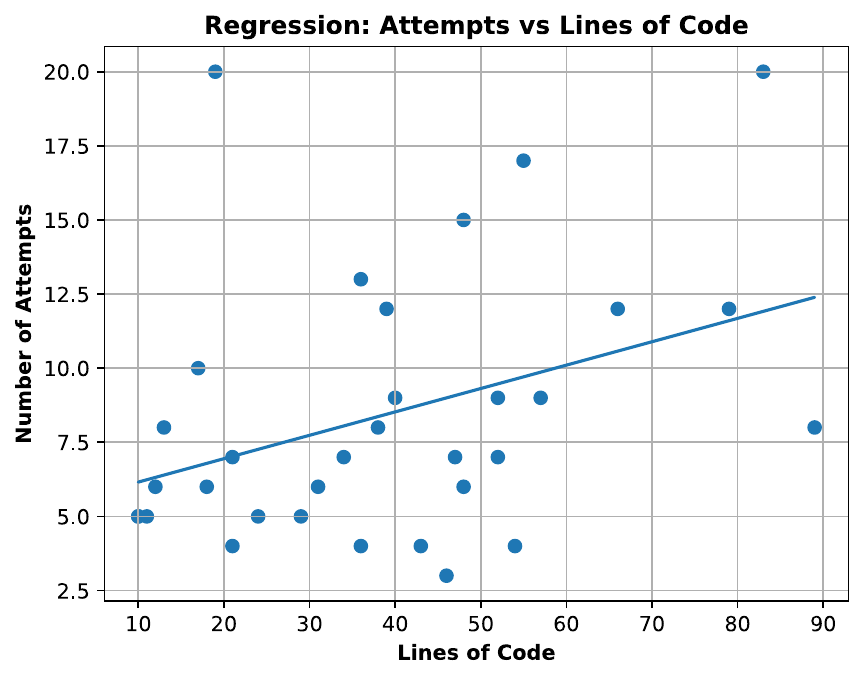}
We conducted a linear regression analysis to examine the relationship between solution size, measured in lines of code (LOC), and the number of attempts required to successfully complete each exercise. Using the combined dataset, the estimated regression coefficients were $\beta_1 \approx 0.079$ for the slope and $\beta_0 \approx 5.37$ for the intercept. The Pearson correlation coefficient was $r \approx 0.37$, with a coefficient of determination of $R^2 \approx 0.14$.

The positive slope indicates that exercises requiring more lines of code tend to require more attempts. Quantitatively, an increase of approximately 10 lines of code is associated with an average increase of about 0.8 attempts.

However, this relationship is moderate rather than strong. The relatively low $R^2$ value indicates that solution length alone explains only about 14\% of the variance in the number of attempts, implying that other factors—such as problem structure, language-specific semantics, and interactions with the type system—play a substantial role in determining model performance.
\begin{table}[h]
\centering
\caption{Classification of Compilation Errors Observed During Baseline Evaluation}
\label{tab:error_classification}
\footnotesize
\setlength{\tabcolsep}{6pt}
\resizebox{\columnwidth}{!}{%
\begin{tabular}{|l|c|}
\hline
\textbf{Error Type} & \textbf{\# of Occurrences} \\
\hline
Undefined name & 123 \\
Ambiguous elaboration & 111 \\
Other / non-compiler errors & 114 \\
Parse / syntax error (general) & 33 \\
Privacy / visibility (not exported / private) & 36 \\
Expected `else' parse error & 22 \\
Missing module / import not found & 15 \\
Type mismatch / unification error & 11 \\
Unknown operator & 2 \\
Entrypoint missing (Main not found) & 1 \\
Totality / termination error & 1 \\
\hline
\end{tabular}}
\end{table}
Table~\ref{tab:error_classification} summarizes the distribution of compilation errors observed during the baseline evaluation of GPT-5 on Idris exercises. The most frequent error categories were undefined names (123 occurrences) and ambiguous elaboration (111 occurrences), indicating that name resolution and scope disambiguation constitute major challenges for the model in a dependently typed language. These errors typically arise from missing helper functions, incorrect imports, or conflicts between locally defined identifiers and standard library definitions, reflecting difficulties in managing Idris’s namespace and elaboration mechanisms.
A substantial number of errors were also classified as non-compiler errors (114 occurrences), which include platform-level messages and execution artifacts rather than true compiler diagnostics. Among genuine compiler errors, parse and syntax errors were relatively common (33 general parse errors and 22 instances of missing \texttt{else} branches), suggesting that the model frequently produces structurally incomplete control-flow constructs in Idris’s \texttt{do}-notation. Privacy and visibility errors (36 occurrences) further indicate challenges in correctly exporting definitions and navigating module boundaries, which are particularly strict in Idris.

Overall, the error distribution highlights that the primary obstacles for GPT-5 in Idris are not deep type-theoretic reasoning failures, but rather surface-level semantic and structural issues related to scope, naming, and compilation discipline. This finding helps explain why Method 4—iterative refinement driven by compiler diagnostics—was particularly effective: compiler-level feedback directly targets the most prevalent error categories, enabling systematic correction of the issues that most frequently prevent successful compilation.

\section{Discussion and Conclusions}

This study confirms that while GPT-5 exhibits strong proficiency in high-resource languages like Python (90\% success) and Erlang (74\%), it struggles significantly with the dependently typed language Idris, achieving a baseline success rate of only 39\%. However, we demonstrate that this gap can be bridged without additional training. By implementing an iterative refinement loop driven by local compiler diagnostics (Method 4), the model's performance improved dramatically to 96\% (54/56 problems). These results directly address our research questions, showing that while general-purpose LLMs lack inherent familiarity with niche syntax, they are highly capable of adapting when guided by precise, error-guided feedback. Although we evaluate only Idris, the core mechanism of Method 4—iterative refinement using precise compiler diagnostics—should generalize to other low-resource, strongly typed, or formal languages (e.g., Agda, Coq, Lean).

The superior performance of compiler-driven refinement compared to document-augmented prompting highlights the nature of the model's difficulties. Our error analysis revealed that the primary obstacles were not deep logical flaws, but surface-level issues such as undefined names and ambiguous elaboration. Generic documentation provided context but failed to correct these specific instance errors. In contrast, the compiler acted as a rigorous verifier, pinpointing exact faults in scope and type definitions. This suggests that for strict functional languages, the compiler provides a far more potent learning signal than natural language prompting or high-level test failures.

% Despite these successes, limitations remain. The weak correlation between lines of code and attempt count ($R^2 \approx 0.14$) suggests that difficulty in Idris may be driven more by type-system interactions than solution length. 

A threat to validity is that some Exercism Idris solutions or closely related code patterns may have appeared in the model's pretraining data, potentially inflating apparent ``new language'' performance.
Future work should test on newly authored or time-held-out Idris tasks (and/or perturbed variants via renaming and rephrased statements) to better distinguish semantic problem solving from recall. Future work should also explore integrating LLMs with formal verification tools extending this compiler-in-the-loop methodology to other formal environments like Coq or Agda.

\bibliographystyle{IEEEtran}
\bibliography{references}

% Generated by IEEEtran.bst, version: 1.14 (2015/08/26)
\begin{thebibliography}{10}
\providecommand{\url}[1]{#1}
\csname url@samestyle\endcsname
\providecommand{\newblock}{\relax}
\providecommand{\bibinfo}[2]{#2}
\providecommand{\BIBentrySTDinterwordspacing}{\spaceskip=0pt\relax}
\providecommand{\BIBentryALTinterwordstretchfactor}{4}
\providecommand{\BIBentryALTinterwordspacing}{\spaceskip=\fontdimen2\font plus
\BIBentryALTinterwordstretchfactor\fontdimen3\font minus \fontdimen4\font\relax}
\providecommand{\BIBforeignlanguage}[2]{{%
\expandafter\ifx\csname l@#1\endcsname\relax
\typeout{** WARNING: IEEEtran.bst: No hyphenation pattern has been}%
\typeout{** loaded for the language `#1'. Using the pattern for}%
\typeout{** the default language instead.}%
\else
\language=\csname l@#1\endcsname
\fi
#2}}
\providecommand{\BIBdecl}{\relax}
\BIBdecl

\bibitem{chen2021codex}
\BIBentryALTinterwordspacing
M.~Chen, J.~Tworek, H.~Jun, Q.~Yuan, H.~P. d.~O. Pinto \emph{et~al.}, ``Evaluating large language models trained on code,'' \emph{arXiv preprint arXiv:2107.03374}, 2021. [Online]. Available: \url{https://arxiv.org/abs/2107.03374}
\BIBentrySTDinterwordspacing

\bibitem{cassano2022multiplE}
\BIBentryALTinterwordspacing
F.~Cassano, A.~Jangda, A.~Guha \emph{et~al.}, ``{MultiPL-E}: A scalable and extensible approach to benchmarking neural code generation,'' \emph{arXiv preprint arXiv:2208.08227}, 2022. [Online]. Available: \url{https://arxiv.org/abs/2208.08227}
\BIBentrySTDinterwordspacing

\bibitem{cassano2023multiplT}
\BIBentryALTinterwordspacing
F.~Cassano, J.~Gouwar, F.~Lucchetti, C.~Schlesinger, A.~Freeman, C.~J. Anderson, M.~Q. Feldman, M.~Greenberg, A.~Jangda, and A.~Guha, ``Knowledge transfer from high-resource to low-resource programming languages for code {LLMs},'' \emph{arXiv preprint arXiv:2308.09895}, 2023. [Online]. Available: \url{https://arxiv.org/abs/2308.09895}
\BIBentrySTDinterwordspacing

\bibitem{madaan2023selfrefine}
\BIBentryALTinterwordspacing
A.~Madaan, N.~Tandon, P.~Gupta, S.~Hallinan, L.~Gao, S.~Wiegreffe, U.~Alon, N.~Dziri, S.~Prabhumoye, Y.~Yang, S.~Gupta, B.~P. Majumder, K.~Hermann, S.~Welleck, A.~Yazdanbakhsh, and P.~Clark, ``Self-refine: Iterative refinement with self-feedback,'' \emph{arXiv preprint arXiv:2303.17651}, 2023. [Online]. Available: \url{https://arxiv.org/abs/2303.17651}
\BIBentrySTDinterwordspacing

\bibitem{chen2024selfdebugging}
X.~Chen, M.~Lin, N.~Sch{\"a}rli, and D.~Zhou, ``Teaching large language models to self-debug,'' in \emph{International Conference on Learning Representations (ICLR)}, 2024.

\bibitem{ravi2025llmloop}
\BIBentryALTinterwordspacing
R.~Ravi, D.~Bradshaw, S.~Ruberto, G.~Jahangirova, and V.~Terragni, ``{LLMLOOP}: Improving {LLM}-generated code and tests through automated iterative feedback loops,'' in \emph{International Conference on Software Maintenance and Evolution (ICSME), Tool Track}, 2025. [Online]. Available: \url{https://valerio-terragni.github.io/assets/pdf/ravi-icsme-2025.pdf}
\BIBentrySTDinterwordspacing

\bibitem{yasunaga2020drrepair}
\BIBentryALTinterwordspacing
M.~Yasunaga and P.~Liang, ``Graph-based, self-supervised program repair from diagnostic feedback,'' in \emph{International Conference on Machine Learning (ICML)}, 2020. [Online]. Available: \url{https://arxiv.org/abs/2005.10636}
\BIBentrySTDinterwordspacing

\bibitem{first2023baldur}
\BIBentryALTinterwordspacing
E.~First, M.~N. Rabe, T.~Ringer, and Y.~Brun, ``Baldur: Whole-proof generation and repair with large language models,'' in \emph{ACM Joint European Software Engineering Conference and Symposium on the Foundations of Software Engineering (ESEC/FSE)}, 2023. [Online]. Available: \url{https://arxiv.org/abs/2303.04910}
\BIBentrySTDinterwordspacing

\bibitem{yang2023leandojo}
\BIBentryALTinterwordspacing
K.~Yang, A.~M. Swope, A.~Gu, R.~Chalamala, P.~Song, S.~Yu, S.~Godil, R.~Prenger, and A.~Anandkumar, ``{LeanDojo}: Theorem proving with retrieval-augmented language models,'' in \emph{NeurIPS Datasets and Benchmarks Track}, 2023. [Online]. Available: \url{https://arxiv.org/abs/2306.15626}
\BIBentrySTDinterwordspacing

\bibitem{blinn2024typedholes}
\BIBentryALTinterwordspacing
A.~Blinn, X.~Li, J.~H. Kim, and C.~Omar, ``Statically contextualizing large language models with typed holes,'' \emph{arXiv preprint arXiv:2409.00921}, 2024. [Online]. Available: \url{https://arxiv.org/abs/2409.00921}
\BIBentrySTDinterwordspacing

\end{thebibliography}

\end{document}